\documentclass{ifacconf}
\usepackage{multirow}
\usepackage{placeins}
\usepackage{graphicx}      
\usepackage{natbib}        
\usepackage{amsmath}
\usepackage{xurl} 

\usepackage{enumerate}
\usepackage{tikz}
\usetikzlibrary{shapes.geometric, arrows}
\usetikzlibrary{shapes,arrows}
\usepackage{float}
\usepackage{xfrac}
\usepackage{graphics} 
\usepackage{tabularx} 
\usepackage{graphicx}
\usepackage{epstopdf}
\usepackage{dblfloatfix}
\usepackage{epsfig} 
\usepackage{mathptmx} 
\usepackage{times} 
\usepackage{amsmath} 
\usepackage{amssymb}  
\usepackage{comment}
\usepackage{caption}
\usepackage{mathtools}
\usepackage{svg}
\DeclarePairedDelimiter\abs{\lvert}{\rvert}
\makeatletter
\let\oldabs\abs
\def\abs{\@ifstar{\oldabs}{\oldabs*}}
\begin{document}
\begin{frontmatter}
\vspace{-15pt}
\title{\small © 2024 the authors. This work has been accepted to IFAC for publication at the 5th IFAC Workshop on Cyber-Physical Human Systems under a Creative Commons Licence CC-BY-NC-ND}
\vspace{-15pt}
\title{Analysis of human steering behavior differences in human-in-control and autonomy-in-control driving} 

\author[First]{Rene E. Mai}
\author[First]{Agung Julius}
\author[First]{Sandipan Mishra}
\address[First]{Rensselaer Polytechnic Institute, 
   Troy, NY 12180 USA (e-mail: mair@rpi.edu, agung@ecse.rpi.edu, mishrs2@rpi.edu).}
\begin{abstract}
Steering models (such as the generalized two-point model) predict human steering behavior well when the human is in direct control of a vehicle. In vehicles under autonomous control, human control inputs are not used; rather, an autonomous controller applies steering and acceleration commands to the vehicle. For example, human steering input may be used for state estimation rather than direct control. We show that human steering behavior changes when the human no longer directly controls the vehicle and the two are instead working in a shared autonomy paradigm. Thus, when a vehicle is not under direct human control, steering models like the generalized two-point model do not predict human steering behavior. We also show that the error between predicted human steering behavior and actual human steering behavior reflects a fundamental difference when the human directly controls the vehicle compared to when the vehicle is autonomously controlled. Moreover, we show that a single distribution describes the error between predicted human steering behavior and actual human steering behavior when the human's steering inputs are used for state estimation and the vehicle is autonomously controlled, indicating there may be a underlying model for human steering behavior under this type of shared autonomous control. Future work includes determining this shared autonomous human steering model and demonstrating its performance.

\end{abstract}
\end{frontmatter}

\section{Introduction}
\vspace{-3pt}
Driving is a deceptively simple task, which humans learn to master with apparent ease; however, humans often overestimate their mastery of driving, along with their ability to multitask \citep{guo_can_2024, held_multitasking_2024}, leading to dangerous accidents. Thus, many hope to replace flawed human drivers with fully autonomous vehicles. However, studies show that relatively common phenomena like poor lane markings can dramatically hamper current assistive technology like lane-keep assist \citep{Peiris2022}. Additionally, future vehicle implementations must be able to withstand false lane markings, which function as physical adversaries to autonomous vehicles \citep{Boloor2019}. Humans are less likely to be deceived by adversarial changes or confusing lane markings, but human behavior is far from simple to replicate, much less interpret \citep{lappi_visuomotor_2018, nash_simulation_2022, mai_generalized_2024}.

Research on shared autonomy addresses multiple human-machine interaction methods, from using humans as examples \citep{fu_learning_2024} to providing personality through driver preferences \citep{karagulle_safe_2024, Matsushita2020}. Methods that fuse human and autonomous control may directly combine the actions based on a heuristic \citep{Zhou2021}, or use full autonomous control while providing the human feedback about how their control inputs differ from the autonomous controller's \citep{Zhang2021}. However, all shared autonomy systems risk losing human trust and being disabled if they perform poorly \citep{Akash2020, Wang2016}.

Likewise, research on modeling human steering behavior takes many forms. One of the most well-known models is the two-point visual control model, which describes human steering as a proportional-integral controller \citep{salvucci_two-point_2004}. Researchers have modified the two-point visual control model to include additional human dynamics such as mixed anticipatory and corrective steering, intermittent attention and vision, and reaction time \citep{lappi_visuomotor_2018}. Others use different models such as linear quadratic regulators on visual angles or data-driven models with diverse sensory inputs \citep{nash_simulation_2022, negash_driver_2023}. Prior work showed that, by expanding the order of the two-point model and including vehicle states as well as visual angles, driver-dependent tuning as described in \citep{ortiz_characterizing_2022} might not be necessary \citep{mai_generalized_2024}. This rich body of research, however, does not address how human behavior changes in a shared autonomy environment, although it is clear that human steering behavior will differ in shared autonomy environments, and likely between different types of shared autonomy \citep{karagulle_safe_2024}.

\begin{figure}
    \centering
    \setlength\belowcaptionskip{-1\baselineskip}
    \includegraphics[width=.4\textwidth]{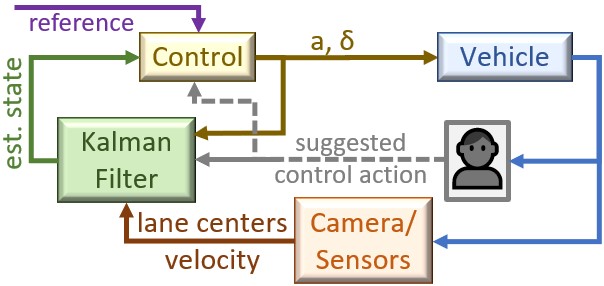}
    \caption{\textit{Human-as-advisor architecture, reproduced from \citep{mai_human-as-advisor_2023}}}
    \label{fig:human-as-advisor}
\end{figure}

\begin{figure}
    \centering
    \includegraphics{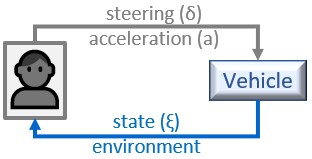}
    \caption{\textit{Typical human-in-control architecture used for identifying human steering models}}
    \label{fig:model identification}
\end{figure}

In this paper, we investigate how human steering behavior changes from a fully human-controlled vehicle to a vehicle controlled via shared autonomy. We refer to the two different control modes as ``human-in-control" and ``autonomy-in-control," respectively. Specifically, we look at how human steering behavior changes from human-in-control to autonomy-in-control when interacting with a vehicle using the human-as-advisor architecture. In human-as-advisor, humans provide suggested control actions; rather than taking the suggested action, however, the state estimator interprets the suggested action through a human model, which allows the state estimator to infer the vehicle's state  \citep{mai_human-as-advisor_2023}. This is illustrated in Fig. \ref{fig:human-as-advisor}. The architecture when the human directly controls a simulated vehicle is shown in Fig. \ref{fig:model identification}. 

Thus, this paper answers the question \textit{does human steering differ significantly when the human is directly controlling a vehicle versus when the human steering input is used to estimate the vehicle state?} We use three methods to answer this question. First,  we analyze the whiteness of the modeling error for human-in-control versus autonomy-in-control trajectories generated by one participant. Second, we look at the distribution of the modeling error for human-in-control and autonomy-in-control trajectories for the same participant. Third, we use the Kolmogorov-Smirnov two-sample test to examine whether the modeling error for the human-in-control and autonomy-in-control trajectories come from the same distribution. In each case, there are striking and consistent differences between the modeling error for the human-in-control and autonomy-in-control steering trajectories. Moreover, the second and third analyses show the autonomy-in-control steering trajectories cluster and have consistent dynamics, which indicates there is likely a \emph{different} model that describes autonomy-in-control human steering behavior.

\section{Vehicle simulation and environment}
\label{sec: vehicle simulation}
We used the Simulink Highway Lane Following Toolbox to simulate the vehicle and environment. The vehicle is nominally controlled using the kinematic bicycle model described in Sec. \ref{sec: vehicle and sensor model}; however, the vehicle simulated includes realistic dynamics such as tire forces and stiffness, aerodynamic drag, and differential forces across the front and rear axles \citep{matlab_bicycle_model}.

\begin{figure}
    \centering
    \includegraphics[width=0.45\textwidth]{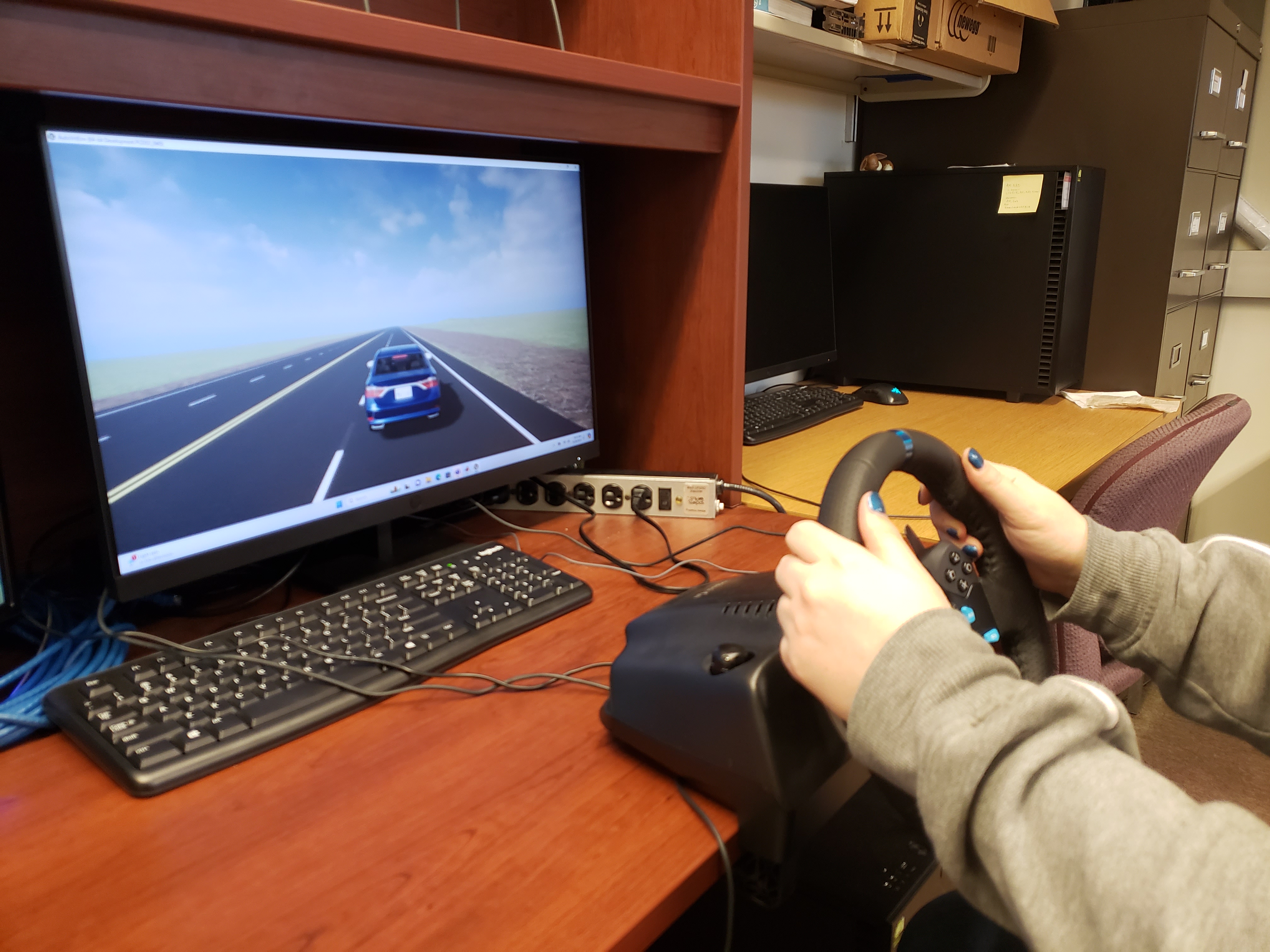}
    \caption{Human data collection setup}
    \label{fig: simulation setup}
\end{figure}

\begin{figure}
    \centering
    \includegraphics[width=0.45\textwidth]{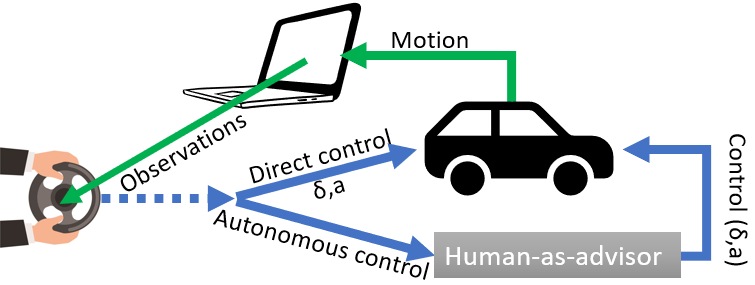}
    \caption{Human-in-the-loop simulation setup and data collection}
    \label{fig: HITL collection}
\end{figure}

The simulation setup is shown in Fig. \ref{fig: simulation setup}.
The third person point of view was chosen to account for peripheral awareness, which the first person point of view from the Highway Lane Following Toolbox (through the windshield only) does not provide. Fig. \ref{fig: HITL collection} illustrates the simulation setup and signal flow used to collect human data.

\section{Mathematical models for control design}
There are numerous autonomous vehicle control strategies. In this section, we describe the development of the human-as-advisor architecture and controller used in this paper.
\subsection{Simplified vehicle and sensor model}
\label{sec: vehicle and sensor model}
\begin{figure}
\centering
\includegraphics[width=.45\textwidth]{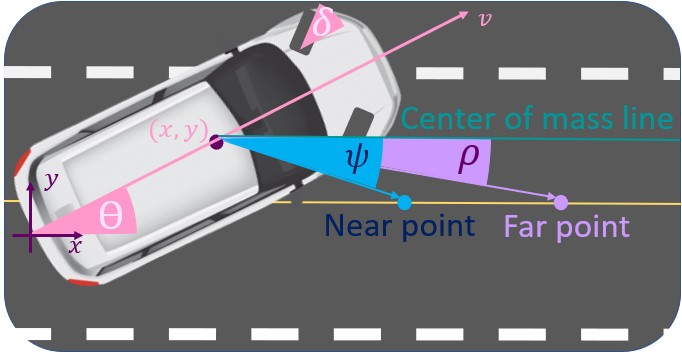}
    \caption{\textit{Near-point and far-point illustration with kinematic bicycle model}}
\vspace{-2mm}
    \label{fig:near-point far-point with kinematic bicycle model}
\end{figure}

\begin{figure}[h]
    \centering
    \includegraphics[width=0.45\textwidth]{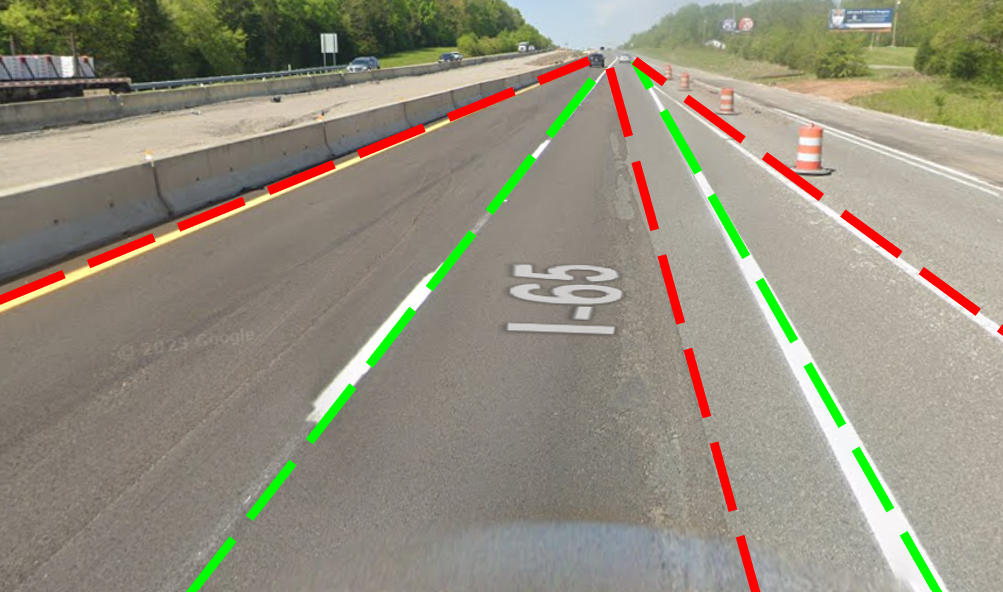}
    \caption{\textit{Ambiguous lane center in a construction zone near Spring Hill, TN (Annotated from Google Maps, retrieved September 2023)}}
    \label{fig: unclear lane center}
\end{figure}

 We use a rear-wheel bicycle kinematic model as shown in Fig. \ref{fig:near-point far-point with kinematic bicycle model} to develop the controller.
The vehicle states used in the bicycle model are
\begin{eqnarray}
\label{eqn_NonlinearContState}
\xi = \begin{bmatrix} \dot{x} \\ \dot{y} \\ \dot{v} \\ \dot{\theta}
\end{bmatrix} = 
\begin{bmatrix} v \cos(\theta) \\ v \sin(\theta) \\ a \\ v \kappa \tan(\delta)
\end{bmatrix} \triangleq \begin{bmatrix} \text{longitudinal velocity} \\ \text{lateral velocity} \\ \text{acceleration} \\ \text{yaw rate} 
\end{bmatrix}.
\end{eqnarray}
For human-in-control scenarios, the driver commands a steering angle $\delta$ and acceleration $a$, which are related to the second derivative of the vehicle's longitudinal and lateral position, $\ddot x$ and $\ddot y$ such that:
\begin{eqnarray}  \label{eqn_FeedbackLin}
\begin{bmatrix}
\ddot x \\
\ddot y%
\end{bmatrix}
\triangleq \underset{\triangleq R(v,\theta )}{\underbrace{%
\begin{bmatrix}
\cos (\theta ) & -v^{2}\kappa \sin (\theta ) \\
\sin (\theta ) & v^{2}\kappa \cos (\theta )%
\end{bmatrix}%
}}%
\begin{bmatrix}
a \\
\tan (\delta )%
\end{bmatrix}.
\end{eqnarray}
The simulated vehicle travels down a straight highway section with multiple lanes and poor lane markings that cause the lane center to be uncertain, as shown in Fig. \ref{fig: unclear lane center}. The possible lane markings in Fig. \ref{fig: unclear lane center} are highlighted in green (the true lane markings) and red (false or ``ghost" lane markings).
We model the uncertain lane center as a bias state, which cannot be controlled and can only be observed by the human. With this bias state defined, we can discretize the system of Eqs. \eqref{eqn_NonlinearContState}-\eqref{eqn_FeedbackLin} with sampling time $T_s$, yielding
\begin{eqnarray}
 \label{eq: disc time closed loop}
\xi(k+1) = \underset{\triangleq A}{\underbrace{\left[ \begin{array}{cccc}
         1 & 0 & 0 & 0\\
         0 & 1 & T_s & 0 \\
         0 & 0 & 1 & 0 \\
         0 & 0 & 0 & 1
     \end{array} \right]}}\xi(k)+\underset{\triangleq B}{\underbrace{\left[ \begin{array}{cc}
         T_s & 0  \\
         0 & \frac{T_s^2}{2} \\
         0 & T_s \\
         0 & 0
     \end{array}\right]}}\underset{\triangleq u}{\underbrace{\begin{bmatrix}
         \ddot x(k) \\ \ddot y(k)
     \end{bmatrix}}}, 
 \end{eqnarray}  
where $ \left[\begin{array}{cccc}
          \xi_1  &
          \xi_2 &
          \xi_3 &
          \xi_4
     \end{array}\right]^T = \left[\begin{array}{cccc}
          \dot x  &
          y &
          \dot y & b 
\end{array}\right]^T$. 
For the human-in-control scenario, we express the human control inputs as $\ddot x$ and $\ddot y$; for autonomy-in-control, we use feedback linearization and thus control $\ddot x$ and $\ddot y$.

In the autonomy-in-control scenario, we formulate a proportional-derivative controller tracking a reference lateral position $y_d(k)$ and longitudinal velocity $\dot x_d(k)$ such that $u_1 = k_{\dot x}\left(\dot x_d(k)-\xi_1(k)\right)$ and $u_2 = k_y\left(y_d(k)-\xi_2(k)\right) + k_{\dot y}\left(\dot y_d(k)-\xi_3(k)\right)$. 

Defining $\xi_d(k) = \begin{bmatrix}
        \dot x_d(k) & y_d(k) & \dot y_d(k)
\end{bmatrix}^T$, the closed-loop autonomy-in-control system becomes 
\begin{equation}
\label{eq: closed-loop autonomy with controller}
    \xi(k+1) = (A-BK)\xi(k)+BK\xi_d(k)
\end{equation}
where $K \triangleq \begin{bmatrix}
        k_{\dot x} & 0 & 0 \\ 0 & k_y & k_{\dot y}
    \end{bmatrix}$. The gains $k_{\dot x}$, $k_{y}$, and $k_{\dot y}$ can be tuned for different performance goals. Here, we set $k_{\dot x} = k_{y} = k_{\dot y} = 1$.

We model the uncertain lane center as a Gaussian mixture model in which each mixture component represents a possible lane center. The $i$th component of the Gaussian mixture model at time step $k$ represents the vehicle's state assuming that the $i$th lane center is the correct lane center, where each component has a likelihood of being true (called the component's weight) and the weights sum to 1. Mathematically, we write this as
\begin{equation}
    \mathcal{G}_i(k) \sim \mathcal{N}(\hat \xi_i(k), \hat \Sigma_i(k), w_i(k)), i \in \{0, 1, 2, ... N\} \text{ where}
\end{equation}
$N$ is the number of components in the Gaussian mixture.

The formulation of Eq. \eqref{eq: closed-loop autonomy with controller} assumes the vehicle state $\xi$ is directly measured; this is of course not possible in practice. Thus, for the autonomy-in-control scenario, the vehicle uses two sensors, a speedometer measuring longitudinal velocity and a camera measuring lateral position in the lane. The sensor model is thus $z_1 = \xi_1+ w_1, \ \ z_2 = \xi_2+ \xi_4 + w_2,$ where $w_{1,2}$ are zero-mean Gaussian measurement noise. These measurements are provided to the state estimator, which is an extended Kalman filter as in \citep{mai_generalized_2024}.

The vehicle is controled through $\delta$ and $a$, which are related to $\ddot x$ and $\ddot y$ by Eq. \eqref{eqn_FeedbackLin}. The vehicle must therefore include accelerometers that measure $\ddot x$ and $\ddot y$, which are the inputs to the plant in \eqref{eq: disc time closed loop}.

\subsection{Generalized two-point human steering model}

The generalized two-point human steering model uses the two visual angles described in \citep{salvucci_two-point_2004}, termed the near-point and far-point angles. The near-point angle, $\phi$, is the angle between the vehicle's current heading and the center of the lane the vehicle is in at the near-point distance, shown on Fig. \ref{fig:near-point far-point with kinematic bicycle model}. We can thus define $\phi \triangleq \theta + \psi$. Likewise, the far-point angle $\Omega$ is the angle between the vehicle's heading and the center of the lane the vehicle is in at the near-point distance, shown on Fig. \ref{fig:near-point far-point with kinematic bicycle model}; thus, we also define $\Omega \triangleq \theta + \rho$. The generalized two-point model uses these visual angles and their history, along with the vehicle's lateral velocity, to predict human steering angles, such that
\begin{multline}
    \label{eq: fitted general model}
    \!\!\delta(k)\!\! =\!\! \sum_{i=1}^2\!\! a_i \delta(k\!-\!i) \!\!+\!\! \sum_{i=0}^3 \!\!b_i \phi(k\!-\!i) \!\!+\!\! c_0 \Omega(k) \!+\! \sum_{i=0}^1\!\! d_i \xi_3(k\!-\!i),
\end{multline}
where $a_0 = 1.47$, $a_1 = 0.51$, $b_0 = -5.73$, $b_1 = 17.32$, $b_2 = -17.65$, $b_3 = 6.12$, $c_0 = 0.11$, $d_0 = 0.02$, and $d_1 = -0.02$ \citep{mai_generalized_2024}. The parameters $a_i$, $b_i$, $c_i$, and $d_i$ were found by minimizing the one-step prediction error.

\section{Steering behavior analysis}
\label{sec: steering analysis methods}
To analyze how well the generalized model in \eqref{eq: fitted general model} predicts human steering behavior, we must first define the prediction error. We define the prediction error at time step $k$ as $\epsilon(k)$, which is the difference between the true steering angle at time $k$, $\delta(k)$, and the one step ahead predicted steering angle based on Eq. \eqref{eq: fitted general model}, $\bar \delta(k)$, such that
\begin{equation}
\label{eq: model error def}
    \epsilon(k) = \bar \delta(k) - \delta(k).
\end{equation}

\subsection{Autocorrelation analysis}
The autocorrelation of a signal--here, the prediction error $\epsilon(k)$--is used to examine whether the signal contains dynamic or periodic information that may be obscured by noise. In a discrete-time system such as in this paper, the error value at each time step, $\epsilon(k)$, is examined for similarity to errors at other time steps. The autocorrelation of white noise replicates the Dirac delta function, indicating that each error value is independent of the others, with some tolerance (generally 0.05). Importantly, a model that fully captures the dynamics of the system it is estimating will have white prediction errors; thus, if the model of Eq. \eqref{eq: fitted general model} represents human steering behavior well, the error in Eq. \eqref{eq: model error def} will be white noise.

Prior work showed the prediction error $\epsilon(k)$ is white for validation data from the participant that supplied the model's training data \citep{mai_generalized_2024}. We also noted the prediction erorr $\epsilon(k)$ is white for several other participants, a surprising result \citep{mai_generalized_2024}. In Sec. \ref{section: data analysis} we again observe that the prediction error is white--but only for the human-in-control data. 

\subsection{Kolmogorov-Smirnoff two-sample test}
In this section we discuss the Kolmogorov-Smirnov two-sample test as described in \citep{1992nrfa.book.....P}. The Kolmogorov-Smirnov two-sample test verifies whether two cumulative distribution functions are likely to be drawn from the same base distribution; we use this test to determine whether the residuals are drawn from the same distribution. Specifically, the Kolmogorov-Smirnov two-sample test looks at the difference between univariate cumulative distribution function (CDF) $C_{1,n}$ and $C_{2,m}$, where 1 and 2 indicate the different distributions being compared and $n$ and $m$ the number of samples in each distribution. The value $\underset{\epsilon}{\text{sup}}\vert C_{1,n}(\epsilon)-C_{2,m}(\epsilon)\vert$ is compared to $D_{n,m}(\alpha)$, where $\alpha$ is the power of the test (the probability of a false positive, typically 0.05). $D_{n,m}(\alpha)$ is defined as
\begin{equation}
\label{eq: ks test statistic def}
D_{n,m}(\alpha) = \sqrt{-\ln\left(\frac{\alpha}{2}\right)\frac{1+\frac{m}{n}}{2m}}.
\end{equation}
If $\underset{\delta}{\text{sup}}\vert C_{1,n}(\epsilon)-C_{2,m}(\epsilon)\vert > D_{n,m}(\alpha)$, then the cumulative distribution functions $C_{1,n}(\epsilon)$ and $C_{2,m}(\epsilon)$ represent two different distributions with confidence $\alpha$. Conversely, if
\begin{equation}
\label{eq: KS test for null hypo}
    \underset{\epsilon}{\text{sup}}\vert C_{1,n}(\epsilon)-C_{2,m}(\epsilon)\vert \leq D_{n,m}(\alpha)
\end{equation}
then $C_{1,n}(\epsilon)$ and $C_{2,m}(\epsilon)$ represent the same distribution with power $\alpha$ (the null hypothesis is true with probability $1-\alpha$).

\section{Experimental data collection}
We collected data using the Simulink Highway Lane Following Toolbox, as described in Sec. \ref{sec: vehicle simulation}. The participant provided 20 driving trajectories, each of which is 30 seconds long, resulting in: (1) 10 human-in-control trajectories, such as those described in \cite{mai_generalized_2024}\footnote{We performed new data collection because prior data was collected over several days; the new data sets were each collected on a single day.} and (2) 10 autonomy-in-control trajectories, where the autonomous controller described in Section \ref{sec: vehicle and sensor model} controls the vehicle and the human-provided steering angles are used for state estimation as shown in Fig. \ref{fig:human-as-advisor}. 

In each test, the vehicle traveled a straight stretch of highway. The participant was directed to attempt to center the vehicle in the right-hand lane, but was not instructed to maintain or change velocity. For both tests, the vehicle's initial state was a two-component Gaussian mixture model with components
\begin{eqnarray}
\vspace{-3pt}
\label{eq: gaussianmixturedef}
\!\!\mathcal{G}_1 \sim \mathcal{N} (\xi_1, I_4, 0.5), 
\mathcal{G}_2 \sim \mathcal{N} (\xi_2, I_4, 0.5), \  \text{where} 
\end{eqnarray} 
\begin{equation*}
    \xi_1 = \!\begin{bmatrix}
    15 & -0.5 & 0 & 0
    \end{bmatrix}^T, \xi_2 = \!\begin{bmatrix}
   15 & 1.3 & 0 & -1.8
    \end{bmatrix}^T.
\end{equation*}
The autonomy-in-control trajectories used the control law in Eq. \eqref{eq: closed-loop autonomy with controller} and an extended Kalman filter as described in \cite{mai_generalized_2024} to estimate the vehicle's state based on the human-suggested steering angles. 

\section{Comparison of steering model errors between human-in-control and autonomy-in-control}
\label{section: data analysis}

\begin{figure}[h]
    \centering
    \includegraphics[width=0.45\textwidth]{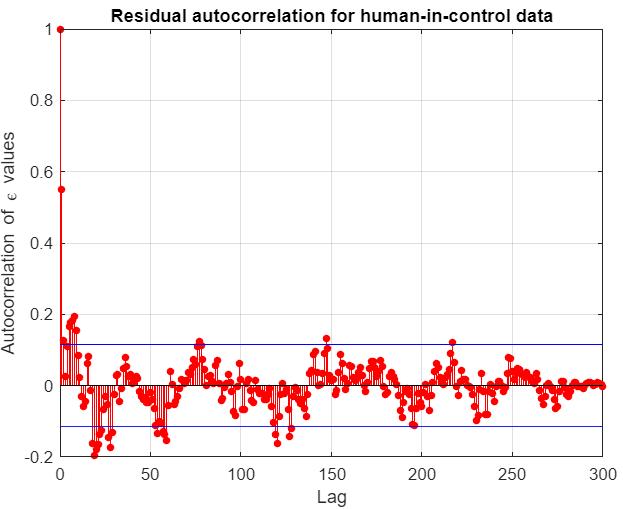}
    \caption{\textit{Autocorrelation of the error between predicted and actual steering angle for a representative human-in-control trajectory.}}
    \label{fig: human in control autocorr}
\end{figure}

As previewed in Sec. \ref{sec: steering analysis methods}, the autocorrelation of $\epsilon$ shows striking differences between human-in-control and autonomy-in-control trajectories. The human-in-control scenario has white residuals across all 10 trajectories, as previous work observed \citep{mai_generalized_2024}. A sample autocorrelation for the human-in-control scenario is shown in Fig. \ref{fig: human in control autocorr}. 

\begin{figure}[h]
    \centering
\includegraphics[width=0.45\textwidth]{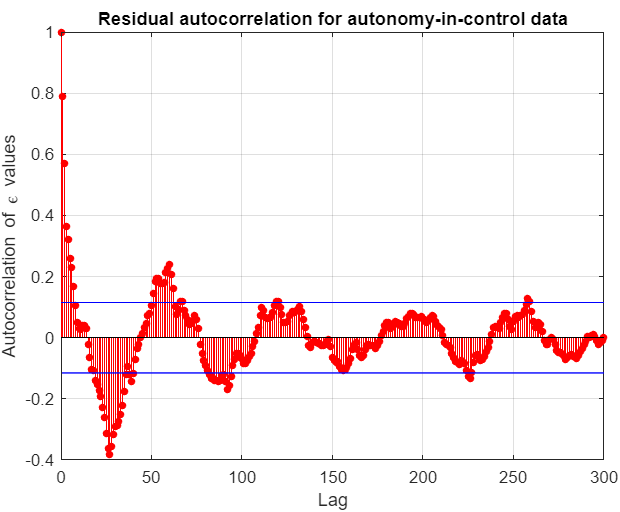}
        \caption{\textit{Autocorrelation of the error between predicted and actual steering angle for a representative autonomy-in-control trajectory. }}
    \label{fig: autonomy in control autocorr}
\end{figure}

In contrast, \textit{not a single one of the 10} autonomy-in-control trajectory has white residuals when the generalized steering model is used to predict \textit{the same participant's} steering behavior; a sample autocorrelation is shown in Fig. \ref{fig: autonomy in control autocorr}. This indicates that the generalized steering model does not accurately capture steering behavior in the autonomy-in-control scenario. 

\begin{figure*}
    \centering
   \includegraphics[width=0.99\textwidth]{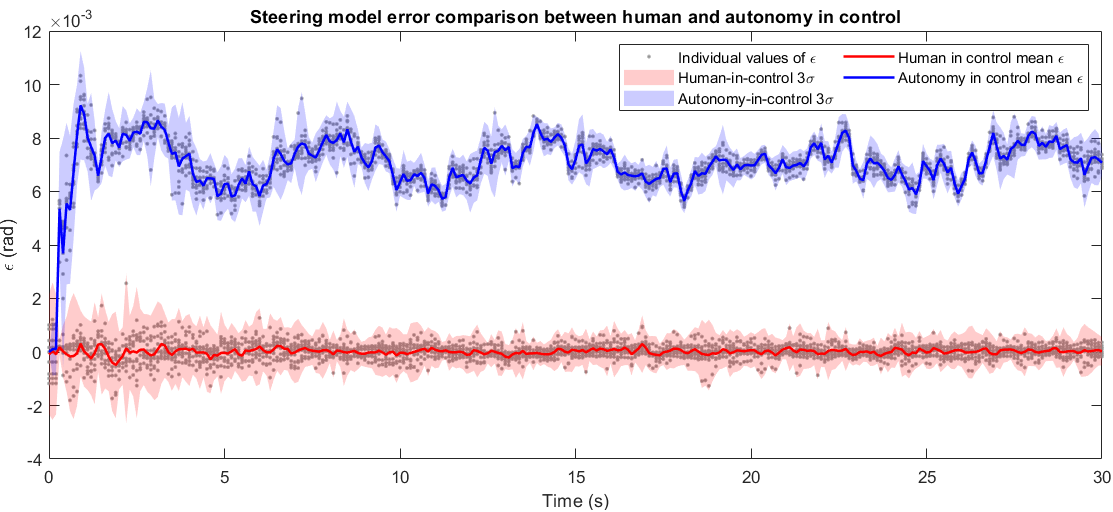}
   \caption{\textit{Error between steering model and true steering values for human-in-control and autonomy-in-control scenarios}}
\vspace{-2mm}
    \label{fig:steering model error comparison}
\end{figure*}
The mean and standard deviation of the residual, $\epsilon$, between the true human steering input and the steering input predicted by the generalized steering model for the human-in-control compared to the autonomy-in-control scenario clearly shows different steering behavior. As Fig. \ref{fig:steering model error comparison} shows, the mean $\epsilon$ for the autonomy-in-control scenario is well over 3 standard deviations (3$\sigma$) away from the mean $\epsilon$ for the human-in-control scenario, and the two error trajectories are close to each other only early in the trajectory (when the autoregressive history is incomplete).

\begin{figure*}
    \centering
\includegraphics[width=0.99\textwidth]{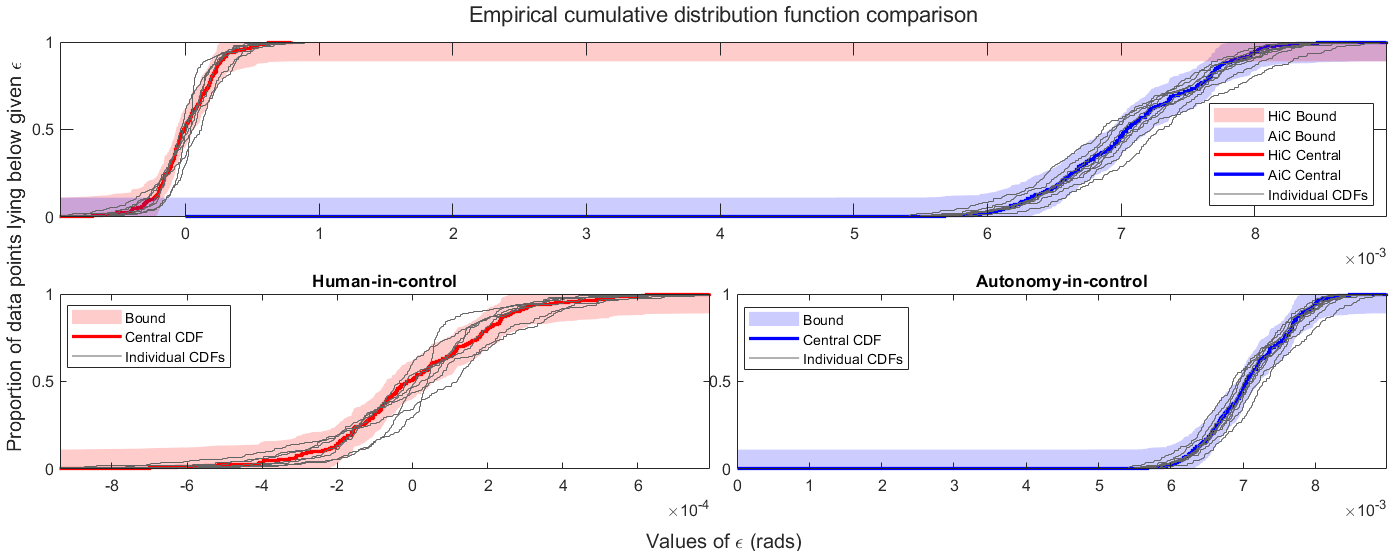}
    \caption{\textit{Comparison of empirical cumulative distribution functions between human-in-control and autonomy-in-control}}
\vspace{-2mm}
    \label{fig:ecdf comparison}
\end{figure*}

Finally, the Kolmogorov-Smirnov two-sample test confirms the empirical evidence that the residual $\epsilon$ in the human-in-control and the autonomy-in-control scenarios are drawn from different distributions. We first formed empirical cumulative distribution functions (CDFs) for all 20 trajectories and found the central distribution for both the human-in-control and autonomy-in-control scenarios. We then used Eq. 
\eqref{eq: KS test for null hypo} to find the upper and lower bounds of a distribution that is not considered significantly different from the central CDF with 95\% significance. The resulting CDFs and bounds are shown in Fig. \ref{fig:ecdf comparison}.
As Fig. \ref{fig:ecdf comparison} makes clear, the CDFs have different central errors and scales, reflecting the different mean error for each scenario. 

Looking at the Kolmogorov-Smirnov test, we find that 3 of the human-in-control CDFs differ from the central human-in-control CDF enough to reject the null hypothesis, while only 2 of the autonomy-in-control CDFs differ enough from the central autonomy-in-control CDF to reject the null hypothesis. If we look at individual error values $\epsilon$, rather than entire CDFs, we find that just 12\% of individual values of $\epsilon$ for the human-in-control distribution and 9.7\% of individual values of $\epsilon$ for the autonomy-in-control distribution fall outside the 95\% confidence level. Thus, it seems likely that the error values for the human-in-control scenario were drawn from the same distribution, while the autonomy-in-control error values were also drawn from a single distribution. 

The Kolmogorov-Smirnov test also makes it abundantly clear that \textit{these two distributions are not the same}; that is, the model for human-in-control steering does not predict steering for autonomy-in-control. In fact, when we cross-test the human-in-control and autonomy-in-control CDFs, the Kolmogorov-Smirnov test rejects the null hypothesis universally with a confidence level of well over 99.999999\% (one in a million). Thus, the two distributions fundamentally differ. More remarkable than the differences between the two distributions, however, is how closely the autonomy-in-control distributions (and, indeed, the values of $\epsilon$ as shown in Fig. \ref{fig:steering model error comparison}) are grouped. This may indicate a different steering model can be used to predict human steering input in the autonomy-in-control scenario--precisely what a shared autonomy implementation like human-as-advisor requires.

\section{Conclusion}
This paper found that human steering behavior differs significantly between human-in-control and autonomy-in-control scenarios. We demonstrated that, although human-in-control steering behavior is well-predicted by the generalized two-point human steering model, this same model does not predict human steering behavior in a shared autonomy architecture. We further demonstrated that the distribution of the steering model error for the autonomy-in-control scenario shows there is likely a central error distribution for autonomy-in-control driving. This suggests that it may be possible to fit a steering model that predicts human steering input in a shared autonomy architecture such as human-as-advisor. Future work includes: (1) fitting a human steering model for shared autonomy such as human-as-advisor, (2) verifying if changing the autonomous controller gains significantly changes human steering behavior, and (3) expanding tests to include more complex tracks such as curves and lane changing.
\section{Acknowledgements}

This work was supported by the 2024-2025 Link Foundation Modeling, Simulation \& Training Fellowship for Rene Mai.

\bibstyle{ifacconf}
\bibliography{ifacconf}

\begin{thebibliography}{20}
\providecommand{\natexlab}[1]{#1}
\providecommand{\url}[1]{\texttt{#1}}
\providecommand{\urlprefix}{URL }
\expandafter\ifx\csname urlstyle\endcsname\relax
  \providecommand{\doi}[1]{doi:\discretionary{}{}{}#1}\else
  \providecommand{\doi}{doi:\discretionary{}{}{}\begingroup \urlstyle{rm}\Url}\fi

\bibitem[{Akash et~al.(2020)Akash, McMahon, Reid, and Jain}]{Akash2020}
Akash, K., McMahon, G., Reid, T., and Jain, N. (2020).
\newblock Human trust-based feedback control: Dynamically varying automation transparency to optimize human-machine interactions.
\newblock \emph{IEEE Control Systems}, 40, 98--116.

\bibitem[{Boloor et~al.(2019)Boloor, He, Gill, Vorobeychik, and Zhang}]{Boloor2019}
Boloor, A., He, X., Gill, C., Vorobeychik, Y., and Zhang, X. (2019).
\newblock Simple physical adversarial examples against end-to-end autonomous driving models.
\newblock \emph{2019 IEEE International Conference on Embedded Software and Systems, ICESS 2019}, 1.

\bibitem[{Fu et~al.(2024)Fu, Zhao, and Finn}]{fu_learning_2024}
Fu, Z., Zhao, T.Z., and Finn, C. (2024).
\newblock Learning {Bimanual} {Mobile} {Manipulation} with {Low}-{Cost} {Whole}-{Body} {Teleoperation}.

\bibitem[{Guo et~al.(2024)Guo, Liu, Lu, and Jing}]{guo_can_2024}
Guo, R., Liu, Y., Lu, H.J., and Jing, A. (2024).
\newblock Can you accurately monitor your behaviors while multitasking? {The} effect of multitasking on metacognition.
\newblock \emph{Psychological Research}, 88(2), 580--593.

\bibitem[{Held et~al.(2024)Held, Rieger, and Borst}]{held_multitasking_2024}
Held, M., Rieger, J.W., and Borst, J.P. (2024).
\newblock Multitasking {While} {Driving}: {Central} {Bottleneck} or {Problem} {State} {Interference}?
\newblock \emph{Human Factors}, 66(5), 1564--1582.
\newblock Publisher: SAGE Publications Inc.

\bibitem[{Karagulle et~al.(2024)Karagulle, Aréchiga, Best, DeCastro, and Ozay}]{karagulle_safe_2024}
Karagulle, R., Aréchiga, N., Best, A., DeCastro, J., and Ozay, N. (2024).
\newblock A {Safe} {Preference} {Learning} {Approach} for {Personalization} {With} {Applications} to {Autonomous} {Vehicles}.
\newblock \emph{IEEE Robotics and Automation Letters}, 9(5), 4226--4233.
\newblock Conference Name: IEEE Robotics and Automation Letters.

\bibitem[{Lappi and Mole(2018)}]{lappi_visuomotor_2018}
Lappi, O. and Mole, C. (2018).
\newblock Visuomotor control, eye movements, and steering: {A} unified approach for incorporating feedback, feedforward, and internal models.
\newblock \emph{Psychological Bulletin}, 144(10), 981--1001.

\bibitem[{Mai et~al.(2023)Mai, Mishra, and Julius}]{mai_human-as-advisor_2023}
Mai, R., Mishra, S., and Julius, A. (2023).
\newblock Human-as-advisor in the loop for autonomous lane-keeping.
\newblock In \emph{2023 {American} {Control} {Conference} ({ACC})}, 3895--3900. IEEE, San Diego, CA, USA.

\bibitem[{Mai et~al.(2024)Mai, Sears, Roessling, Julius, and Mishra}]{mai_generalized_2024}
Mai, R., Sears, K., Roessling, G., Julius, A., and Mishra, S. (2024).
\newblock Generalized two-point visual control model of human steering for accurate state estimation.
\newblock \urlprefix\url{http://arxiv.org/abs/2406.03622}.

\bibitem[{Mathworks(2024)}]{matlab_bicycle_model}
Mathworks (2024).
\newblock Implement a single track {3DOF} rigid vehicle body to calculate longitudinal, lateral, and yaw motion - {Simulink}.
\newblock \urlprefix\url{https://www.mathworks.com/help/driving/ref/bicyclemodel.html}.

\bibitem[{Matsushita et~al.(2020)Matsushita, Sato, Sakura, Sawada, Shin, and Inoue}]{Matsushita2020}
Matsushita, H., Sato, K., Sakura, M., Sawada, K., Shin, S., and Inoue, M. (2020).
\newblock Rear-wheel steering control reflecting driver personality via human-in-the-loop system.
\newblock volume 2020-Octob, 356--362. Institute of Electrical and Electronics Engineers Inc.

\bibitem[{Nash and Cole(2022)}]{nash_simulation_2022}
Nash, C.J. and Cole, D.J. (2022).
\newblock A {Simulation} {Study} of {Human} {Sensory} {Dynamics} and {Driver}–{Vehicle} {Response}.
\newblock \emph{Journal of Dynamic Systems, Measurement, and Control}, 144(061002).

\bibitem[{Negash and Yang(2023)}]{negash_driver_2023}
Negash, N.M. and Yang, J. (2023).
\newblock Driver {Behavior} {Modeling} {Toward} {Autonomous} {Vehicles}: {Comprehensive} {Review}.
\newblock \emph{IEEE Access}, 11, 22788--22821.

\bibitem[{Ortiz et~al.(2022)Ortiz, Thorpe, Perez, Luster, Pitts, and Oishi}]{ortiz_characterizing_2022}
Ortiz, K.R., Thorpe, A.J., Perez, A., Luster, M., Pitts, B.J., and Oishi, M. (2022).
\newblock Characterizing {Within}-{Driver} {Variability} in {Driving} {Dynamics} {During} {Obstacle} {Avoidance} {Maneuvers}.
\newblock \emph{IFAC-PapersOnLine}, 55(41), 13--19.

\bibitem[{Peiris et~al.(2022)Peiris, Newstead, Berecki-Gisolf, Chen, and Fildes}]{Peiris2022}
Peiris, S., Newstead, S., Berecki-Gisolf, J., Chen, B., and Fildes, B. (2022).
\newblock Quantifying the lost safety benefits of adas technologies due to inadequate supporting road infrastructure.
\newblock \emph{Sustainability (Switzerland)}, 14, 1--23.

\bibitem[{{Press} et~al.(1992){Press}, {Teukolsky}, {Vetterling}, and {Flannery}}]{1992nrfa.book.....P}
{Press}, W.H., {Teukolsky}, S.A., {Vetterling}, W.T., and {Flannery}, B.P. (1992).
\newblock \emph{{Numerical recipes in FORTRAN. The art of scientific computing}}, 614--622.
\newblock Cambridge University Press, second edition.

\bibitem[{Salvucci and Gray(2004)}]{salvucci_two-point_2004}
Salvucci, D.D. and Gray, R. (2004).
\newblock A {Two}-{Point} {Visual} {Control} {Model} of {Steering}.
\newblock \emph{Perception}, 33(10), 1233--1248.

\bibitem[{Wang et~al.(2016)Wang, Pynadath, and Hill}]{Wang2016}
Wang, N., Pynadath, D.V., and Hill, S.G. (2016).
\newblock Trust calibration within a human-robot team: Comparing automatically generated explanations.
\newblock \emph{ACM/IEEE International Conference on Human-Robot Interaction}, 2016-April, 109--116.

\bibitem[{Zhang et~al.(2021)Zhang, Tron, and Khurshid}]{Zhang2021}
Zhang, D., Tron, R., and Khurshid, R.P. (2021).
\newblock Haptic feedback improves human-robot agreement and user satisfaction in shared-autonomy teleoperation.
\newblock \emph{Proceedings - IEEE International Conference on Robotics and Automation}, 2021-May, 3306--3312.

\bibitem[{Zhou et~al.(2021)Zhou, Chai, Yin, and Shi}]{Zhou2021}
Zhou, Z., Chai, C., Yin, W., and Shi, X. (2021).
\newblock Developing and evaluating an human-automation shared control takeover strategy based on human-in-the-loop driving simulation.

\end{thebibliography}
                                                   







\end{document}